\documentclass[twocolumn,showpacs,nofootinbib,aps,10pt,prd]{revtex4}
\usepackage{graphicx,dcolumn,bm,amsfonts,amsmath,color,xcolor,amsthm,multirow}
\usepackage[colorlinks=true, citecolor=blue, linkcolor=blue,urlcolor=blue]{hyperref}
\usepackage{slashed}
\usepackage{tensor}
\usepackage{ulem}
\usepackage{setspace}
\usepackage{cancel}
\usepackage{txfonts}
\DeclareMathAlphabet{\mathcal}{OMS}{cmsy}{m}{n}
\DeclareSymbolFont{largesymbols}{OMX}{cmex}{m}{n}

\newcommand{\mysection}[1]{%
 \par\medskip
 \hspace*{0.2em}\textit{#1}.--\ignorespaces
}
\allowdisplaybreaks

\begin{document}
\title{Quasi-four-body semileptonic decay $\bar B^0\to a_0(1450)^+ (\to\pi^+ \eta)\ell^-\bar\nu_\ell$ within $a_0(1450)$-resonance leading-twist LCDA under QCDSR}

\author{Ya-Lin Song}
\address{Department of Physics, Guizhou Minzu University, Guiyang 550025, P.R.China}
\author{Yin-Long Yang}
\address{Department of Physics, Guizhou Minzu University, Guiyang 550025, P.R.China}
\author{Yan-Ting Yang}
\address{Department of Physics, Guizhou Minzu University, Guiyang 550025, P.R.China}
\author{Dong Huang}
\address{Department of Physics, Guizhou Minzu University, Guiyang 550025, P.R.China}
\author{Hai-Bing Fu}
\email{fuhb@gzmu.edu.cn}
\address{Department of Physics, Guizhou Minzu University, Guiyang 550025, P.R.China}

\begin{abstract}
The finite width of broad scalar resonances should be taken into account when relating semileptonic $B$-meson transitions to experimentally reconstructable multihadron final states. The quasi-four-body semileptonic decays $\bar B^0\to a_0(1450)^+ (\to\pi \eta)\ell^-\bar\nu_\ell$ with $\ell=(e,\mu,\tau)$ provide a platform for investigating the scalar $a_0(1450)^+$-resonance as the lowest-Fock-state quark-antiquark component. We calculate the first five odd $\xi$-moments with $n=(1,3,5,7,9)$ of the $a_0(1450)$-resonance twist-2 light-cone distribution amplitude (LCDA) using QCD sum rules within the background-field theory, including vacuum condensates up to dimension six. Based on these moments, two twist-2 LCDA schemes are constructed: a light-cone harmonic oscillator model (S1), whose free parameters are fitted to the first five odd moments, and a truncated Gegenbauer expansion (S2) retaining only $a_1(\mu)$ and $a_3(\mu)$.
 The $\bar{B}^0 \to a_0(1450)^+$ transition form factors are then calculated using light-cone sum rules. At large recoil region, we obtain: $f_+^{\rm(S1)}(0)=0.404^{+0.048}_{-0.054}$,
 $f_-^{\rm(S1)}(0)=-0.403^{+0.047}_{-0.052}$,
 $f_+^{\rm(S2)}(0)=0.398^{+0.047}_{-0.053}$,
 $f_-^{\rm(S2)}(0)=-0.404^{+0.053}_{-0.048}$,
 $f_{\rm T}^{\rm(S1)}(0)=0.501^{+0.067}_{-0.060}$,
 $f_{\rm T}^{\rm(S2)}(0)=0.516^{+0.069}_{-0.062}$.
 The finite-width effect of $a_0(1450)$-resonance is further incorporated through a two-channel Flatt\'e parametrization including the $\pi \eta$ and $K\bar{K}$ channels. The $\pi \eta$ invariant-mass distribution exhibits a pronounced resonance peak around $s\simeq m_{a_0(1450)}^2$. The spectra of the electron and muon channels are nearly identical, while the $\tau$-channel is significantly suppressed because the larger lepton mass reduces the available phase space. After integration over the full quasi-four-body phase space, the branching fractions for $\ell=(e,\mu,\tau)$ are $(0.883^{+0.224}_{-0.221},0.881^{+0.225}_{-0.221},
 0.408^{+0.164}_{-0.145})\times 10^{-4}$ for the S1 scheme, and $(0.783^{+0.199}_{-0.196},0.781^{+0.199}_{-0.196},
 0.353^{+0.144}_{-0.124})\times 10^{-4}$ for the S2 scheme. We expect that these results will provide useful guidance for future experimental and theoretical studies of $\bar B^0\to a_0(1450)^+ (\to\pi \eta)\ell^-\bar\nu_\ell$ decay and help improve our understanding of the internal structure of $a_0(1450)$.
\end{abstract}
\maketitle

\mysection{Introduction}\label {Sec:I}
In semileptonic $B$-meson decays involving broad scalar resonances, treating the intermediate resonance as an on-shell particle of fixed mass is generally insufficient to describe the experimentally reconstructed multihadron final states. The reason is that the propagation of a broad resonance depends not only on the resonance mass and total width but also on different strong decay channels together with their phase space factors, which leads to a pronounced $s$-dependence of the resonance line shape and of the dihadron invariant-mass distribution. The $a_0(1450)$ is a typical broad scalar resonance, with its mass and total width approximately given by $1.44~\rm{GeV}$ and $258~\rm{MeV}$~\cite{ParticleDataGroup:2026mpi}. Its observed strong decay channels include $\pi\eta$, $K\bar{K}$ and $\pi\eta'$. Among them, the $\pi\eta$ and $K\bar{K}$ channels are particularly relevant for studying the $a_0(1450)$-resonance properties. As early as 1994, the Crystal Barrel Collaboration observed the $a_0(1450)$ in the $S$-wave $\pi\eta$ system and subsequently identified the corresponding resonance contribution in the $K\bar{K}$ final state~\cite{Bugg:2008ig, CrystalBarrel:1994arw}. In searches for the decays $B^0 \to a_0(1450)^- \pi^+$ and $B^0 \to a_0(1450)^- K^+$, the BABAR Collaboration also reconstructed the $a_0(1450)^-$ via the $\eta \pi^-$ channel and set upper limits on the product branching fractions of $2.3\times 10^{-6}$ and $3.1\times 10^{-6}$ at the $90\%$ confidence level~\cite{BaBar:2007yhw}. Therefore, for semileptonic decays involving the $a_0(1450)$-resonance, a more direct description should focus on the experimentally reconstructable $\pi\eta$ final state rather than treating the $a_0(1450)$ merely as a particle of fixed mass.

Especially for the quasi-four-body semileptonic decay $\bar{B}^0\to a_0(1450)^+ (\to\pi^+ \eta)\ell^-\bar\nu_\ell$ with $\ell=e,\mu,\tau$, one can define $s=(p_\pi +p_\eta)^2$ and $q^2=(p_\ell +p_{\bar{\nu}})^2$, where $s$ and $q^2$ denote the squared invariant masses of the $\pi\eta$ and $\ell \bar{\nu}_\ell$ systems, respectively. In contrast to the fixed-mass three-body treatment, $s$ varies continuously over the allowed quasi-four-body phase space, so the finite width and the resonance line shape of the $a_0(1450)$-resonance enter directly into the $\pi\eta$ invariant-mass distribution. The decay dynamics can therefore be described in terms of the double-differential decay width $d^2\Gamma/dsdq^2$. The $s$-dependence is particularly sensitive to the finite width and the resonance line shape of the $a_0(1450)$, whereas the $q^2$-dependence probes the $\bar B^0\to a_0(1450)^+$ weak transition; the two variables are also correlated through the quasi-four-body phase space. To incorporate the $\pi\eta$ and $K\bar{K}$ channels associated with the $\pi\eta$ final state within a coupled-channel framework, we adopt a two-channel Flatt\'e-type parametrization for the intermediate $a_0(1450)$-resonance. Since there are currently no experimental measurements or theoretical studies of the $\bar{B}^0\to a_0(1450)^+ (\to\pi^+ \eta)\ell^-\bar\nu_\ell$ quasi-four-body semileptonic decay, a theoretical analysis of its double-differential decay width and branching fraction is well motivated.

The internal structure of $a_0(1450)$ resonance has not yet been fully established. If regarded as a physical hadronic state, its full wave function may in general contain conventional $q\bar{q}$, tetraquark $qq\bar{q}\bar{q}$, meson-meson, and other higher Fock components~\cite{Brito:2004tv, Klempt:2007cp}. In the present quasi-four-body process, however, the $a_0(1450)$ is not treated as a stable asymptotic final-state particle, but rather as an intermediate scalar resonance that is produced through the weak $\bar{B}^0 \to a_0(1450)^+$ transition and subsequently decays into the $\pi\eta$ final state. For the description of the weak-production subprocess, the nonperturbative dynamics can be organized in terms of the light-cone Fock expansion. At leading twist, the LCDA characterizes the longitudinal momentum distribution of the valence partons in the lowest two-parton Fock component, while configurations involving additional partons belong to higher Fock components~\cite{Balitsky:1989ry, Chernyak:1990ag}. Such a lowest-Fock-state treatment is commonly adopted in QCDSR and LCSR analyses of scalar-meson LCDAs and heavy-to-light $B\to S$ transition form factors (TFFs)~\cite{Han:2013zg, Wang:2008da}. In particular, previous QCDSR studies have employed the valence-quark component as the dominant contribution in constructing the scalar-meson LCDAs and obtained results compatible with the observed $a_0(1450)$ state~\cite{Han:2013zg}. Accordingly, for the weak $\bar{B}^0 \to a_0(1450)^+$ transition considered here, we retain the lowest two-parton Fock component and describe it through the corresponding LCDAs. This working approximation does not imply that the full physical $a_0(1450)$-resonance is a pure $q\bar{q}$ state, nor does it exclude possible tetraquark, meson-meson, or other higher Fock components. The finite width and coupled-channel dynamics of the intermediate $a_0(1450)$-resonance are treated separately through its resonance line shape in the subsequent quasi-two-body framework.

Within the above quasi-two-body framework, the weak transition part of the quasi-four-body decay amplitude is described by the hadronic matrix element of $\bar B^0\to a_0(1450)^+$, which can be parameterized by the TFFs. The intermediate $a_0(1450)$-resonance subsequently decays into $\pi^+ \eta$ through the strong interaction. Therefore, a reliable determination of the $\bar B^0\to a_0(1450)^+$ TFFs is essential for calculating the quasi-four-body decay width and branching fraction. These TFFs have been studied using perturbative QCD (pQCD), three-point QCD sum rules (3PSR), relativistic quark models (RQM), covariant light-front quark models (CLFQM), and light-cone sum rules (LCSR)~\cite{Song:2025mfm, Wang:2008da, Han:2023pgf, Balitsky:1989ry, Chernyak:1990ag, Huang:2021owr, Galkin:2025emi, Verma:2011yw}. One advantage of the LCSR approach is that the correlation function can be expanded near the light cone $x^2 \simeq 0$ using the operator product expansion (OPE), with long-distance nonperturbative dynamics parameterized by light-cone distribution amplitudes (LCDAs) of different twists. This approach can thus handle perturbatively calculable short-distance contributions and soft contributions determined by the final-state hadron structure within a unified framework, and it is particularly suitable for studies of heavy-to-light transitions in the low and intermediate $q^2$-regions~\cite{Shifman:2001ck, Khodjamirian:2000ds, Duplancic:2008ix, Wang:2007fs, Ali:1993vd, Aliev:1995zlh, Khodjamirian:2000mi, Khodjamirian:2002pk, Khodjamirian:2003eq, Khodjamirian:2005wn, Belyaev:1994zk}. Based on this treatment, we adopt the quasi-two-body approximation in the quasi-four-body decay calculation and take the TFFs to satisfy $f_i(q^2,s)\simeq f_i(q^2)$; that is, the dependence of the TFFs on the $\pi\eta$ invariant-mass squared $s$ is neglected. The $s$-dependence of $a_0(1450)$-resonance is described by the Flatt\'e spectral function $P_{\pi\eta}(s)$. The $\pi\eta$ and $K\bar{K}$ channels enter the resonance propagator through their $s$-dependent two-body phase-space factors. These contributions determine the $a_0(1450)$-resonance line shape and the $\pi\eta$ invariant mass distribution.

In the LCSR framework, the LCDAs of the $a_0(1450)$-resonance provide the essential nonperturbative information for determining the $\bar B^0\to a_0(1450)^+$ TFFs, with the twist-2 LCDA playing a particularly important role. The $a_0(1450)$-resonance twist-2 LCDA $\phi_{2;a_0(1450)}(x,\mu)$ is usually expressed as a Gegenbauer polynomial expansion, whose Gegenbauer moments $a_n(\mu)$ can be determined by the corresponding $\xi$-moments~\cite{Wang:2008da}. These moments characterize the longitudinal momentum distribution of the valence quarks and therefore provide direct constraints on the shape of the LCDA. In the approximate isospin symmetry limit, the even moments are strongly suppressed, whereas the odd moments dominate, leading to a twist-2 LCDA $\phi_{2;a_0(1450)}(x,\mu)$ that is antisymmetric about $x=1/2$. Previous studies have mainly focused on the lowest odd moments, which provide only limited information on the detailed $x$-dependence of the LCDA. Our earlier studies on the twist-2 LCDAs of the pion indicate that relying only on a small number of low order $\xi$-moments and a finite-order truncated Gegenbauer expansion is often insufficient to achieve an accurate description of the LCDA. Higher odd $\xi$-moments are useful for imposing additional constraints on its longitudinal momentum distribution~\cite{Zhong:2022lmn, Huang:2022xny}. To this end, we calculate the first five odd $\xi$-moments of the twist-2 LCDA for the $a_0(1450)$-resonance with $n=(1,3,5,7,9)$ using QCD sum rules (QCDSR) within the background-field theory (BFT) framework, and include contributions from nonperturbative vacuum condensates up to dimension six in the operator product expansion (OPE)~\cite{Hu:2023pdl, Hu:2021lkl, Zhong:2014jla, Zhong:2014fma}.

According to QCD conformal symmetry, the $a_0(1450)$-resonance twist-2 LCDA can generally be expanded in terms of Gegenbauer polynomials, where the Gegenbauer moments contain nonperturbative information on the quark momentum distribution inside the hadron. Since finite-order Gegenbauer expansions are mainly determined by low-order moments, while higher-order moments can provide additional information on the LCDA behavior, incorporating more $\xi$-moments can further constrain the LCDA and help explore the impact of different parametrizations on the predictions of the relevant observables. Based on the calculated odd $\xi$-moments, we construct two twist-2 LCDA parametrizations. One is the light-cone harmonic oscillator (LCHO) model (S1) based on the Brodsky-Huang-Lepage (BHL) prescription, where the model parameters are constrained by the five odd $\xi$-moments~\cite{Huang:2022xny, Yang:2005bv, Huang:1994dy}. The other is a truncated Gegenbauer expansion (S2) that retains only the lowest nonvanishing Gegenbauer moments $a_{1,3}(\mu)$~\cite{Cheng:2005nb}. By comparing different LCDA parametrizations, we investigate the impact of the nonperturbative structure description of the $a_0(1450)$-resonance on the $\bar B^0\to a_0(1450)^+$ TFFs and the subsequent predictions for the quasi-four-body semileptonic decay. In addition, the contributions from the two-particle twist-3 LCDAs are included in the TFF calculation.

\mysection{Theoretical Framework}\label {sec:II}
The LCDAs of the $a_0(1450)$-resonance constitute the essential nonperturbative inputs for the LCSR calculation of the $\bar{B}^0 \to a_0(1450)^+$ TFFs. We therefore begin by determining the $\xi$-moments of its twist-2 LCDA, which are subsequently used to construct the two LCDA schemes adopted in this work. Since the twist-2 moments cannot be fixed by a normalization condition alone, we also introduce the zeroth moment of the two-particle twist-3 $p$-type LCDA, which is used in the determination of the continuum threshold $s_{a_0(1450)}$ and in the extraction of the twist-2 moments~\cite{Ball:2005vx, Lu:2006fr}.
\begin{align}
&\hspace{-0.5cm}\langle 0|\bar{q}_1( z ) \Gamma q_2( -z ) |a_0( 1450 ) ( p ) \rangle
\nonumber\\
&\hspace{-0.5cm}=
\begin{cases}
p_\mu\bar{f}_{a_0( 1450 )}\displaystyle\int_0^1{d}xe^{i\xi (p\cdot z)}\phi _{2;a_0( 1450 )}( x,\mu ) ,&	\Gamma =\gamma_\mu
\\
m_{a_0( 1450 )}\bar{f}_{a_0( 1450 )}\displaystyle\int_0^1{d}xe^{i\xi (p\cdot z)}\phi _{3;a_0( 1450 )}^{p}( x,\mu ), &	\Gamma =\mathbb{I}
\end{cases}
\label{DA defined}
\end{align}
where $\xi = (2x-1)$, $z^2=0$, $f_{a_0(1450)}$ and $m_{a_0(1450)}$ are the decay constant and mass of the $a_0(1450)$-resonance, respectively, while $\bar{f}_{a_0(1450)}$ denotes the scale-dependent decay constant. The corresponding $n$th moment of twist-2 LCDA is defined as $\langle \xi^n_{2;a_0(1450)} \rangle |_\mu = \int_0^1 dx \xi^n \phi_{2;a_0(1450)}(x,\mu)$. In the isospin symmetry approximation, the $a_0(1450)$-resonance twist-2 LCDA is antisymmetric under $x\leftrightarrow (1-x)$, so that the odd moments provide the relevant nonvanishing contributions. Therefore, we calculate the first five odd-order moments with $n=(1,3,5,7,9)$ in this work. Expanding the nonlocal matrix elements in Eq.~(\ref{DA defined}) around the light-cone near $z=0$, the $\xi$-moments can be related to matrix elements of local operators as:
\begin{align}
\langle 0|\bar{q}_1(0) \slashed{z} (iz \cdot & \overleftrightarrow{D})^n q_2(0)| a_0(1450)(p)\rangle
\nonumber \\
&=(z \cdot p)^{n+1} \bar{f}_{a_0(1450)} \langle \xi^n_{2;a_0(1450)} \rangle|_\mu,
\nonumber \\
\langle 0|\bar{q}_1(0) q_2(0)&| a_0(1450)(p)\rangle
\nonumber \\
&= m_{a_0(1450)} \bar{f}_{a_0(1450)} \langle \xi^{p,0}_{3;a_0(1450)} \rangle|_\mu,
\end{align}
Here, $\langle\xi_{3;a_0(1450)}^{p,0}\rangle|_\mu$ is the zeroth-order $\xi$-moment of the two-particle twist-3 $p$-type LCDA of the $a_0(1450)$-resonance. The covariant derivative operator satisfies relation $(iz \cdot \overleftrightarrow D)^n =(iz \cdot \overrightarrow D -iz \cdot \overleftarrow D)^n$.

Following the standard procedure for calculating the twist-2 LCDA moments $\langle \xi^n_{2;a_0(1450)} \rangle|_\mu$ within the QCDSR framework, we introduce the correlation function:
\begin{align}
\Pi_{2;a_0(1450)}^{(n,0)}(z,q) &= i \int d^4x\, e^{iq\cdot x}
\langle 0|T\bigl\{J_n(x), \hat{J}_0^{\dagger}(0)\bigr\}|0\rangle
\nonumber \\
& =(z\cdot q)^{n+1} I_{2;a_0(1450)}(q^2),
\label{correlator}
\end{align}
with the interpolating currents $J_n(x) =\bar{q}_1(x) \slashed{z}(iz \cdot \overleftrightarrow{D})^n q_2(x)$ and $J_0^{\dagger}(0) =\bar{q}_1(0) q_2(0)$. Within the BFT framework, we perform the OPE for the correlator in the deep Euclidean region $(q^2 \ll 0)$. Here, the quantum fields describe short-range perturbative effects, while the classical background fields characterize long-range nonperturbative effects such as vacuum condensates. The nonperturbative contributions of condensates with dimensions higher than six to the $\xi$-moments are extremely small and can be safely neglected. Therefore, the OPE result $I^{\rm{QCD}}_{2;a_0(1450)}(q^2)$ is retained up to dimension six, including contributions from perturbative terms, quark condensate, quark-gluon mixed condensate, two-gluon condensate, three-gluon condensate and four-quark condensate. The vertex operator $(iz \cdot \overleftrightarrow{D})^n$ corresponds to higher-dimensional operators or higher-twist terms. In practical calculations, the $\xi$-moments computed with this operator are almost numerically identical to those obtained with it retained. Thus, the contribution of the vertex operator is neglected in this paper. On the other hand, in the physical region, by inserting a complete set of hadronic states with the same $J^P$-quantum numbers as $a_0(1450)$-resonance into the correlator~(\ref{correlator}), the hadronic expression is obtained as:
\begin{align}
&\mathrm{Im}I_{2;a_0(1450)}^{\rm{had}}(s) =\pi\delta m_{a_0(1450)}\bar{f}_{a_0(1450)}^2(s-m_{a_0(1450)}^2)\langle\xi^{n}_{2;a_0(1450)}\rangle|_\mu
\nonumber \\
& \qquad \times  \langle\xi^{p;0}_{3;a_0(1450)}\rangle|_\mu
 +\mathrm{Im}I^{\mathrm{pert}}_{2;a_0(1450)}(s)
\theta(s-s_{a_0(1450)}),
\end{align}
where $m_q$ denotes the $u$ and $d$ light quark mass. $s_{a_0(1450)}$ is the continuum threshold parameter. Then, according to quark-hadron duality, we combine the above OPE result $I^{\rm{QCD}}_{2;a_0(1450)}(q^2)$ with the hadronic representation via the dispersion relation and perform the following Borel transformation: $\frac{1}{\pi}\frac{1}{M^2}\int dse^{-s/M^2} \mathrm{Im}I^{\rm{had}}_{2;a_0(1450)}(s)= \hat{B}_MI_{2;a_0(1450)}^{\rm{QCD}}(q^2)$. We finally obtain the sum rule for the twist-2 LCDA moments of the $a_0(1450)$-resonance:
\begin{align}
& \langle \xi _{2;a_0(1450)}^n\rangle \big|_\mu \langle \xi _{3;a_0(1450)}^{p;0}\rangle \big|_\mu
= \frac{e^{m_{a_0(1450)}^2/M^2}}{m_{a_0(1450)} \bar f_{a_0(1450)}^2}\bigg\{ \frac1\pi
\nonumber\\
& \qquad \times \int_{m_s^2}^{s_{a_0(1450)}}ds e^{-s/M^2} \mathrm{Im} I^{\mathrm{pert}}(s)
+ \frac{\langle \bar dd\rangle}{M^2} - (-1)^n\frac{\langle \bar uu\rangle}{M^2}
\nonumber\\
& \qquad + \frac{2(n + 3)}{81M^4}\Big[\langle g_s\bar dd\rangle^2 m_u - (-1)^n\langle g_s\bar uu\rangle^2 m_d\Big]
\nonumber\\
& \qquad - \frac{n}{2M^2}\Big[\langle g_s\bar d\sigma TGd\rangle - n (-1)^n\langle g_s\bar u\sigma TGu\rangle \Big]
\nonumber\\
& \qquad + \hat{\mathcal B}_M \hat I_{\langle G^2\rangle}(M^2)
+ \hat{\mathcal B}_M^2 \hat I_{\langle G^3\rangle}(M^2)
+ \hat{\mathcal B}_M \hat I_{\langle q^4\rangle}(M^2)\bigg\},
\label{xi0xin}
\end{align}
where $\langle \bar u u\rangle$ and $\langle \bar{d}d\rangle$ are the quark condensates, $\langle g_s \bar u \sigma TGu\rangle$ and $\langle g_s \bar{d}\sigma TGd\rangle$ are the quark-gluon mixed condensates, $\langle \alpha_s G^2\rangle$ and $\langle g_s^3 fG^3\rangle$ are the two-gluon and three-gluon condensates, respectively. $\langle g_s^2 \bar u u\rangle^2$ and $\langle g_s^2 \bar{d}d\rangle^2$ are the four-quark condensates. For convenience, the explicit expressions for $\mathrm{Im}I^{\mathrm{pert}}(s)$, $\hat{\mathcal{B}}_M\hat{I}_{\langle G^2\rangle}(M^2)$, $\hat{\mathcal{B}}_M\hat{I}_{\langle G^3\rangle}(M^2)$ and $\hat{\mathcal{B}}_M\hat{I}_{\langle q^4\rangle}(M^2)$ are collected in the Appendix. In particular, since $\langle \xi^{p;0}_{3;a_0(1450)} \rangle|_\mu$ depends on the Borel parameter $M^2$, the sum rule~(\ref{xi0xin}) is not used to extract the twist-2 $\xi$-moment directly. Instead, we first determine the product $\langle \xi^n_{2;a_0(1450)}\rangle|_\mu \langle \xi^{p;0}_{3;a_0(1450)} \rangle|_\mu$ from Eq.(\ref{xi0xin}) and independently calculate the zeroth twist-3 moment. Following the same QCDSR procedure, we can obtain the expression for the $0$-th moment $\langle \xi^{p;0}_{3;a_0(1450)} \rangle|_\mu$ of the $a_0(1450)$-resonance twist-3 LCDA,
\begin{align}
&\hspace{-2ex}\langle \xi _{3;a_0(1450)}^{p;0}\rangle \big|_\mu ^2 = \frac{e^{m_{a_0(1450)}^2/M^2}}{m_{a_0(1450)} \bar f_{a_0(1450)}^2}\bigg\{ \int_{m_s^2}^{s_{a_0(1450)}} ds\, e^{-s/M^2} \frac{3M^2}{8\pi^3}
\nonumber\\
&+ \frac{\langle \alpha_s G^2\rangle }{8\pi M^2}  + \frac{\langle \bar dd\rangle}{M^2} \left(\frac{m_d}{2} + m_u\right) +  \frac{\langle\bar uu\rangle}{M^2} \left(\frac{m_u}{2} + m_d\right)
\nonumber\\
&+ \frac{1}{2M^2}\big(m_d\langle g_s \bar d\sigma TGd\rangle  + m_u\langle g_s \bar u\sigma TGu\rangle \big)- \frac{(2 + \kappa^2)}{11664\pi^2}
\nonumber\\
&\times  \big( 144\gamma_E + 936\big) \langle g_s^2\bar qq\rangle + \frac{\langle g_s\bar qq\rangle ^2}{27M^2}\left(\frac{m_d^2}{M^2} - 8\right)\bigg\} ,
\label{xi02}
\end{align}
Finally, dividing Eq.~\eqref{xi0xin} by the square root of Eq.~\eqref{xi02} yields the final expression for the moments. This ratio treatment can eliminate systematic errors arising from sources such as the continuum state, the absence of higher-dimensional condensates, and the selection and determination of the various input parameters~\cite{Zhong:2021epq,Zhong:2022ecl}. Furthermore, since the twist-2 LCDA of $a_0(1450)$-resonance is dominated by the odd-order Gegenbauer terms, its $\xi$-moments can be used to determine the corresponding Gegenbauer moments $a_n(\mu)$ at the same scale.

With the resulting $\xi$-moments at hand, we can further construct the twist-2 LCDA of the $a_0(1450)$-resonance. As an important nonperturbative input for the heavy-to-light TFFs, $\phi_{2;a_0(1450)}(x,\mu)$ describes the longitudinal momentum distribution of the valence quarks in the lowest Fock state. We adopt two different LCDA schemes to examine the model dependence of the twist-2 LCDA on the subsequent predictions for physical observables in this work. For the first scheme (S1), we adopt the LCHO model based on the BHL prescription. The leading-twist light-cone wave function of the $a_0(1450)$-resonance is constructed from the spin-space and spatial wave functions
$\Psi_{2;a_0(1450)}(x,\mathbf{k}_\perp)=\sum_{\lambda_1\lambda_2}
\chi_{a_0(1450)}^{\lambda_1\lambda_2}(x,\mathbf{k}_\perp)
\Psi_{a_0(1450)}^R (x,\mathbf{k}_\perp)$, where $\sum_{\lambda_1\lambda_2} \chi_{a_0(1450)}^{\lambda_1\lambda_2}(x,\mathbf{k}_\perp)$ and $\Psi_{2;a_0(1450)}^R(x,\mathbf{k}_\perp)$ explicit forms can be found in our earlier work~\cite{Song:2025mfm}. After integrating over the transverse momentum $\mathbf{k}_\perp$, the twist-2 LCDA in the S1 scheme is obtained as:
\begin{align}
&\hspace{-2ex}\phi_{2;a_0(1450)}^{\rm(S1)} (x,\mu)= \frac{A_{2;a_0(1450)} \hat{m}_q \beta_{2;a_0(1450)}}{4 \sqrt{2} \pi^{3/2}} \sqrt{x \bar{x}}\varphi_{2;a_0(1450)}(x)
\nonumber \\
& \times \bigg\{\mathrm{Erf}\bigg[\sqrt{\frac{\hat{m}_q^2 + \mu^2}{8\beta_{2;a_0(1450)}^2 x\bar{x}}}\bigg]  -\mathrm{Erf}\bigg[\sqrt{\frac{\hat{m}_q^2}
{8\beta_{2;a_0(1450)}^2 x\bar{x}}}\bigg]\bigg\},
\end{align}
where $\bar{x}=1-x$, the longitudinal distribution function $\varphi_{2;a_0(1450)}(x)=(x\bar{x})^{\alpha_{2;a_0(1450)}}
C^{3/2}_1(\xi)$, $\hat{m}_q =250~\rm{MeV}$ denotes the constituent quark mass. The three free parameters $A_{2;a_0(1450)}$, $\beta_{2;a_0(1450)}$, $\alpha_{2;a_0(1450)}$ can be determined by fitting the first five odd-order $a_0(1450)$-resonance $\xi$-moments with least-squares method.

To compare the differences among the different parametrization schemes for the twist-2 LCDA of the $a_0(1450)$-resonance, we further introduce the second scheme (S2). Based on conformal symmetry, the twist-2 LCDA of the $a_0(1450)$-resonance can also be expressed as a series expansion in Gegenbauer polynomials. In order to provide a simple truncated benchmark and to avoid introducing poorly constrained higher Gegenbauer coefficients, we retain the two lowest nonvanishing odd moments
\begin{align}
\phi_{2;a_0(1450)}^{\rm (S2)}(x,\mu) = 6x\bar{x}[a_1(\mu)C_1^{3/2}(\xi)+ a_3(\mu)C_3^{3/2}(\xi)].
\label{DA truncated form}
\end{align}
The Gegenbauer moments $a_{1,3}(\mu)$ can be obtained from the $a_0(1450)$-resonance twist-2 LCDA $\xi$-moments. Meanwhile, to obtain more accurate TFFs for the $\bar{B}^0 \to a_0(1450)^+$ transition, we also consider the twist-3 LCDAs within the Gegenbauer series expansion, $\phi_{3;a_0(1450)}^{p,\sigma}(x,\mu)$ up to fourth order.

Within the LCSR framework, the $\bar{B}^0 \to a_0(1450)^+$ TFFs are defined by the correlation function (correlator) of the weak current and the interpolating current of the $B$-meson. For this purpose, we adopt the conventional interpolating current method starting from the following correlator:
\begin{align}
\hspace{-2ex}\Pi_\mu^{\rm (I, II)} (p,q) &=i\!\int \! d^4x e^{iq\cdot x}\langle a_0(1450)(p)|{\rm T} \{j_{2 \mu}^{\rm (I, II)}(x),j_{1}(0)\}|0\rangle,
\label{correlator TFFs}
\end{align}
where the first current is $j_1(0) = m_b\bar b(0)i\gamma_5 q_1(0)$. The second current $j_{2\mu}(x)$ has two different forms. When $j_{2\mu}^{\rm I}(x) = \bar{q}_2(x)\gamma_\mu\gamma_5b(x)$ is taken, the correlator reads $\Pi_\mu^{\rm I} (p,q) = F( q^2,(p+q)^2 )p_\mu +\tilde F (p^2, (p+q)^2) q_\mu$, which highlights the vector TFFs $f_{\pm}(q^2)$. When $j_{2\mu}^{\rm II}(x) = \bar{q}_2(x)\sigma_{\mu\nu}\gamma_5q^\nu b(x)$ is taken, the correlator takes the form $\Pi_\mu^{\rm II} (p,q)=F^{\rm T} ( q^2,(p+q)^2)[ p_\mu q^2 -q_\mu (p\cdot q)]$, which highlights the tensor TFF $f_{\rm T}(q^2)$. The light quarks are $q_1 =d$ and $q_2 =u$. On the hadronic side, a complete set of states with the $\bar{B}^0$-meson quantum numbers is inserted into the correlators~(\ref{correlator TFFs}), and the ground-state $\bar{B}^0$ pole is isolated. The heavy-to-light TFFs $f_{\pm,{\rm T}}(q^2)$ can be related to the transition matrix element. On the QCD side, the correlators are expanded near the light cone in terms of the $a_0(1450)$-resonance LCDAs. Matching the two representations through the dispersion relation, invoking quark-hadron duality for the continuum contribution, and performing the Borel transformation in the initial meson momentum squared, we obtain the LCSR expressions for the TFFs, which are also given in the Appendix.

Since the $a_0(1450)$-resonance is a broad unstable resonance with a finite width, we can adopt the intermediate-resonance approximation to describe the quasi-four-body semileptonic decay. In this treatment, the decay amplitude is factorized into the weak transition $\bar{B}^0 \to a_0(1450)^+$ and the subsequent strong decay $a_0(1450)^+ \to \pi^+ \eta$~\cite{Wang:2016wpc},
\begin{align}
\mathcal{A}(\bar{B}^0 \to \pi^+ \eta \ell^-\bar{\nu}_\ell) = \hat{A} \bigg[\frac{i}{D_{a_0(1450)}(s)} ig^{\rm{str}}_{\pi\eta} \bigg],
\end{align}
where $g^{\rm{str}}_{\pi\eta}$ is the strong coupling constant at the $a_0(1450)^+ \to \pi^+ \eta$ vertex, and $\hat{A}$ denotes the weak decay amplitude of the intermediate $a_0(1450)$-resonance, whose hadronic matrix element is parametrized by the TFFs $f_{\pm}(q^2)$ introduced above. Here, $s=(p_\pi +p_\eta)^2$ and $q^2 =(p_\ell +p_{\bar{\nu}})^2$ denote the squared invariant masses of the $\pi\eta$ and $\ell \bar{\nu}_\ell$ systems, respectively. Taking into account the self-energy corrections, finite-width effects, and the couplings of $a_0(1450)$-resonance to its hadronic decay channels, the denominator of the resonance propagator reads~\cite{Achasov:2005hm}:
\begin{align}
D_{a_0(1450)}(s)&=m_{a_0(1450)}^2 -s
\nonumber\\
&+\sum_{M_1M_2} \big[ \operatorname{Re}\Pi_{M_1M_2}(m_{a_0(1450)}^2) - \Pi_{M_1M_2}(s) \big],
\end{align}
where $M_1M_2=(\pi\eta, K\bar{K})$, $\Pi_{M_1M_2}(s)$ denotes the self-energy contribution from the intermediate two hadron state $M_1M_2$. The real part of the self-energy shifts the resonance pole and can be absorbed into the definition of the physical mass $m_{a_0(1450)}$. Once the corresponding strong-decay channel is open, the self-energy develops an imaginary part, which determines the mass-dependent partial width and describes the coupling of the $a_0(1450)$-state: ${\rm Im} \Pi_{M_1M_2}(s) =\sqrt{s} \Gamma_{M_1M_2}(s) = |g_{M_1M_2}^{\rm{str}}|^2/(16\pi) \rho_{M_1M_2}(s)$, where $\rho_{M_1M_2}(s)$ represents the two-body phase-space factor, defined as
\begin{align}
\hspace{-0.2cm}\rho_{M_1M_2}(s) &= \sqrt{\big[1-\frac{(m_{M_1} -m_{M_2})^2}{s} \big]\big[ 1-\frac{(m_{M_1} +m_{M_2})^2}{s}\big]},
\end{align}
for the $a_0(1450)$, we consider the $\pi\eta$ and $K\bar{K}$ channels. After absorbing the real self-energy correction into the physical resonance mass and defining $G_{M_1M_2}=|g_{M_1M_2}^{\rm{str}}|^2/(16\pi)$, the two-channel Flatt\'e denominator becomes:
\begin{align}
D_{a_0}(s) = m_{a_0(1450)}^2 -s -i\big[ G_{\pi\eta}\rho_{\pi\eta}(s) + G_{K\bar{K}}\rho_{K\bar{K}}(s) \big],
\end{align}
the strong coupling constant $g_{M_1M_2}^{\rm{str}}$ can be determined from the corresponding partial decay width,
\begin{align}
\hspace{-0.2cm}|g_{M_1M_2}^{\rm{str}}|=\sqrt{\frac{8\pi m_{a_0(1450)}^2 \Gamma_{a_0(1450)}\mathcal{B}(a_0(1450) \to M_1M_2)}{p_{M_1M_2}}},
\end{align}
where $p_{M_1M_2}=[(m_{a_0(1450)}^2 -(m_{M_1} +m_{M_2})^2)(m_{a_0(1450)}^2 -(m_{M_1} -m_{M_2})^2)]^{1/2}/(2m_{a_0(1450)})$ is the three-momentum of the $\pi$ or $\eta$ in the rest frame of the $\pi \eta$ system. Based on this, the spectral function of $a_0(1450)$-resonance for $\pi^+ \eta$ final state reads:
\begin{align}
P_{\pi\eta}(s) = \frac{G_{\pi\eta}\rho_{\pi\eta}(s)}{\big| m_{a_0(1450)}^2 \!-\! s \!-\! i\big[ G_{\pi\eta}\rho_{\pi\eta}(s) + G_{K\bar{K}}\rho_{K\bar{K}}(s) \big] \big|^2}.
\end{align}
Within the quasi-two-body approximation, we neglect the additional $s$-dependence of the weak transition form factors and take $f_{\pm}(q^2, s) \simeq f_{\pm}(q^2)$; combining the weak transition with the Flatt\'e spectral function and integrating over the lepton angles, the double-differential decay width is given by:
\begin{align}
&\hspace{-2ex}\frac{d^2\Gamma(\bar B^0 \to a_0(1450)^+(\to \pi^+\eta)\ell^-\bar\nu_\ell)}{ds\, dq^2}
= \frac{G_F^2|V_{ub}|^2 m_{\bar B^0}^3}{192\pi^3}
\nonumber\\
&\times\lambda^{3/2}(m_{\bar B^0}^2,s,q^2)
\left(1 - \frac{m_\ell^2}{q^2}\right)^2
\bigg[\left(1 + \frac{m_\ell^2}{2q^2}\right)\big|f_+^{(\mathrm{S1},\mathrm{S2})}(q^2)\big|^2
\nonumber\\
&+ \frac{3m_\ell^2}{2q^2}\frac{(m_{\bar B^0}^2 - s)^2}{\lambda(m_{\bar B^0}^2,s,q^2)}
\bigg|f_+^{(\mathrm{S1},\mathrm{S2})}(q^2) + \frac{q^2}{m_{\bar B^0}^2 - s}f_-^{\rm (S1,S2)}(q^2)\bigg|^2\bigg]
\nonumber\\
&\times \frac{P_{\pi\eta}(s)}{\pi},
\label{formula decay width}
\end{align}
where $\lambda(x, y, z)=x^2 +y^2 +z^2 -2xy -2xz -2yz$ is the K\"all\'en function. Here $s$, as the continuous squared invariant mass of the $\pi\eta$ system, enters directly into $\lambda(m_{\bar{B}^0}^2, s, q^2)$, so that the finite-width effect of the $a_0(1450)$-resonance and the kinematics of the quasi-four-body final state are treated simultaneously. The physical phase space for the quasi-four-body process satisfies $(m_\pi+m_\eta)^2 \leq s \leq (m_{\bar{B}^0}-m_\ell)^2$ and $m_\ell^2 \leq q^2 \leq (m_{\bar{B}^0}-\sqrt{s})^2$. Therefore, the total decay width is obtained by integrating the double-differential rate over the full physical region~\cite{Dosch:2002rh}:
\begin{align}
\Gamma =\int_{(m_\pi+m_\eta)^2}^{(m_{\bar{B}^0}-m_\ell)^2} ds \int_{m_\ell^2}^{(m_{\bar{B}^0}-\sqrt{s})^2} dq^2 \frac{d^2\Gamma}{ds\,dq^2}
\label{eq:total_width}
\end{align}

\begin{figure}[t]
\begin{center}
\includegraphics[width=0.45\textwidth]{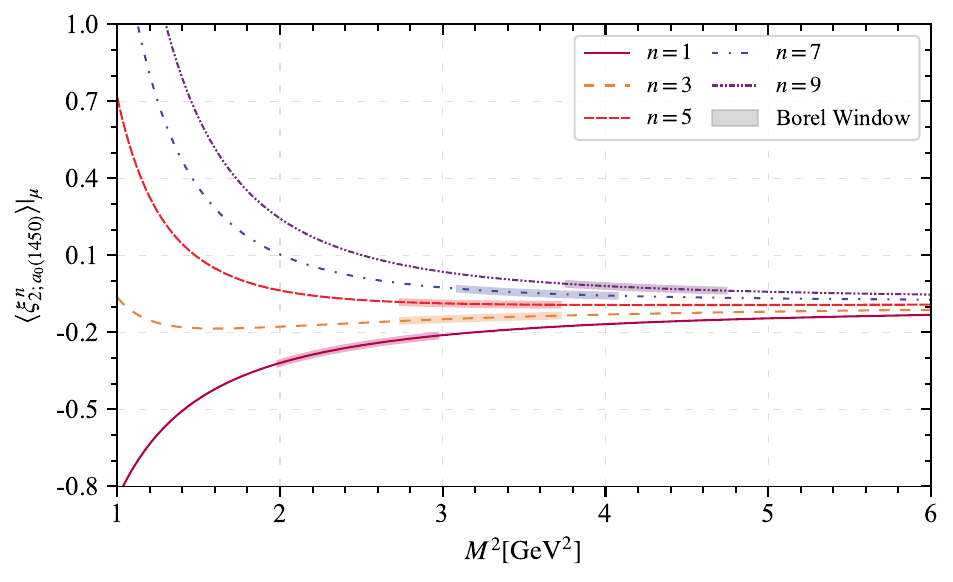}
\end{center}
\caption{The dependence of $\xi$-moments $\langle\xi^n_{2;a_0(1450)}\rangle|_\mu (n=1,3,5,7,9) $ for leading-twist LCDA of $a_0(1450)$-resonance on the Borel Window $M^2$, with all input parameters taken at their central values; the shaded bands indicate the chosen Borel windows.}
\label{Fig:n-moment}
\end{figure}

\mysection{Numerical Analysis}\label {sec:III}
In dealing with the subsequent numerical analysis, we adopt the basic input parameters from  PDG~\cite{ParticleDataGroup:2024cfk}. The relevant hadron masses are $m_{B^0} =5279.72\pm{0.08}~\mathrm{MeV}$, $m_{a_0(1450)}=1439\pm{34}~\mathrm{MeV}$, while the quark masses are $m_b(\bar{m}_b) =4.183\pm{0.007}~\mathrm{GeV}$, $m_u(2\rm{GeV}) =2.16\pm{0.07}~\mathrm{MeV}$ and $m_d(2\rm{GeV}) =4.7\pm{0.07}~\mathrm{MeV}$. The decay constants are taken as $f_{\bar{B}^0} =0.207^{+0.017}_{-0.009}~\mathrm{GeV}$ and $f_{a_0(1450)}=0.46\pm{0.05}~\mathrm{GeV}$ at the initial scale $\mu_0 =1~\rm{GeV}$~\cite{Gelhausen:2013wia, Cheng:2005nb}. The renormalization scale for the QCDSR calculation of the twist-2 LCDA moments is taken as $\mu=\sqrt{M^2}$. The continuum threshold $s_{a_0(1450)}=8.1~{\rm GeV^2}$ is determined from the normalization of the zeroth moment of the twist-3 $p$-type LCDA. The nonperturbative vacuum condensates up to dimension six are taken as~\cite{Zhong:2014jla, Zhong:2021epq, Colangelo:2000dp}:
\begin{align}
\langle \bar{q}q \rangle ~(2~\rm{GeV})~&=~(-2.417^{+0.227}_{-0.114})\times 10^{-2}~\rm{GeV}^3,
\nonumber \\
\langle g_s\bar{q}\sigma TGq \rangle (2~\rm{GeV})~&=~(-1.934^{+0.188}_{-0.103})\times 10^{-2} ~\rm{GeV}^5,
\nonumber \\
\langle g_s \bar{q}q \rangle^2 (2~\rm{GeV})~&=~(2.082^{+0.734}_{-0.697}) \times10^{-3}~\rm{GeV}^6,
\nonumber \\
\langle g^2_s \bar{q}q \rangle^2 ~&=~(7.420^{+2.614}_{-2.483})\times 10^{-3}~\rm{GeV}^6,
\nonumber \\
\langle \alpha_s G^2 \rangle ~&=~0.038\pm{0.011}~\rm{GeV}^4,
\nonumber \\
\langle g^3_{s} fG^3 \rangle ~&\simeq~0.045~\rm{GeV}^6,
\nonumber \\
\kappa ~&=~0.74\pm{0.03}.
\end{align}

Meanwhile, the typical energy scale $\mu_k =(m_{B^0}^2 -m_b^2)^{1/2} \simeq 3 ~\mathrm{GeV}$ and evolve the scale-dependent input parameters using the standard renormalization group equations (RGE) are used for the $B$-meson decay processes~\cite{Zhong:2021epq}. The Borel windows for $\xi$-moments are determined by requiring both OPE convergence and sufficient suppression of continuum contributions. For the moments with $n=(1,3,5,7,9)$, the continuum contributions are below $15\%,~25\%,~25\%,~30\%$ and $40\%$, respectively, while the dimension-six contributions are restricted to less than $5\%$. In addition, the extracted moments should exhibit weak dependence on $M^2$ within the selected windows. Fig.~\ref{Fig:n-moment} shows the $M^2$ dependence of the first five odd moments, $\langle \xi^n_{2;a_0(1450)} \rangle|_\mu$ with $n=(1,3,5,7,9)$. The moments change dramatically for $M^2 \in[1,2]~\rm{GeV}^2$ but become increasingly stable for $M^2 > 2~\rm{GeV}^2$, and their absolute values decrease overall with increasing $n$. The shaded regions denote the Borel windows selected from the continuum and OPE convergence constraints.

\begin{table}
\footnotesize
\centering
\caption{The first five odd $a_0(1450)$-resonance twist-2 LCDA moments and Gegenbauer moments at $\mu_{0,k}$ scale.}
\label{tab:moments}
\begin{tabular}{l l l l l}
\hline
$n$~~~~ & $\langle\xi^n_{2;a_0(1450)} \rangle|_{\mu_0}$ ~~~ & $a_n(\mu_0)$ ~~~~~~~~~~~~~~~ & $\langle\xi^n_{2;a_0(1450)} \rangle|_{\mu_k}$ ~~~& $a_n(\mu_k)$ \\
\hline
$1$ & $-0.345^{+0.086}_{-0.086}$ & $-0.576^{+0.143}_{-0.142}$ & $-0.207^{+0.051}_{-0.051}$ & $-0.345^{+0.086}_{-0.085}$ \\
$3$ & $-0.208^{+0.030}_{-0.029}$ & $-0.315^{+0.035}_{-0.038}$ & $-0.117^{+0.018}_{-0.018}$ & $-0.150^{+0.026}_{-0.024}$ \\
$5$ & $-0.123^{+0.018}_{-0.018}$ & $+0.238^{+0.078}_{-0.082}$ & $-0.070^{+0.011}_{-0.011}$ & $+0.098^{+0.017}_{-0.015}$ \\
$7$ & $-0.058^{+0.026}_{-0.026}$ & $+1.227^{+0.748}_{-0.761}$ & $-0.037^{+0.008}_{-0.008}$ & $+0.460^{+0.060}_{-0.064}$ \\
$9$ & $-0.039^{+0.023}_{-0.023}$ & $-3.584^{+1.858}_{-1.917}$ & $-0.025^{+0.012}_{-0.012}$ & $-1.245^{+1.154}_{-1.137}$ \\
\hline
\end{tabular}
\end{table}

The resulting $\xi$-moments and Gegenbauer moments at $\mu_{0,k}$ are listed in Table~\ref{tab:moments}. Compared with the predictions of other QCDSR studies, $\langle\xi^1 \rangle =-0.35\pm{0.07}$, $a_1 =-0.58\pm{0.12}$, $\langle\xi^3 \rangle =-0.24\pm{0.06}$, and $a_3 =-0.49\pm{0.15}$ at $\mu_0$~\cite{Cheng:2005nb}, our results for $\langle\xi^1_{2;a_0(1450)} \rangle|_{\mu_0}$ and $a_1(\mu_0)$ are consistent with these values, while $\langle\xi^3_{2;a_0(1450)} \rangle|_{\mu_0}$ and $a_3(\mu_0)$ are lower than the QCDSR values by $0.03$ and $0.17$, respectively, which may be attributed to the different treatment of $s_{a_0(1450)}$. In addition, we determine the higher odd moments with $n=5,7,9$, which provide more complete constraints on the twist-2 LCDA of the $a_0(1450)$-resonance. We then determine the two twist-2 LCDA schemes. For $\phi_{2;a_0(1450)}^{\rm(S1)}(x,\mu)$, we perform a least-squares fit of the three free parameters $A_{2;a_0(1450)}$, $\beta_{2;a_0(1450)}$ and $\alpha_{2;a_0(1450)}$ of the LCHO model to the five odd $\xi$-moments with $n=(1,3,5,7,9)$. The optimal fitted parameters at the initial scale $\mu_0$ are $A_{2;a_0(1450)} = -50$, $\alpha_{2;a_0(1450)} = -2.7$, and $\beta_{2;a_0(1450)} = 0.2$, with goodness of fit $X_{\rm{min}}^2 = 0.574455$ and $P_{\chi^2_{\rm min}} = 0.902255$, and $A_{2;a_0(1450)} = -106$, $\alpha_{2;a_0(1450)} = -1.3$, and $\beta_{2;a_0(1450)} = 0.3$, with goodness of fit $X_{\rm{min}}^2 = 0.483199$ and $P_{\chi^2_{\rm min}} = 0.922567$, at $\mu_k$. The fitting probabilities exceed $90\%$ at both $\mu_0$ and $\mu_k$. For $\phi_{2;a_0(1450)}^{\rm(S2)}(x,\mu)$, we adopt $a_1(\mu_0)=-0.576^{+0.143}_{-0.142}$ and $a_3(\mu_0)=-0.315^{+0.035}_{-0.038}$ from Table~\ref{tab:moments} and retain only these two lowest nonvanishing odd terms in the Gegenbauer expansion to construct the truncated twist-2 LCDA.

\begin{figure}
\begin{center}
\includegraphics[width=0.45\textwidth]{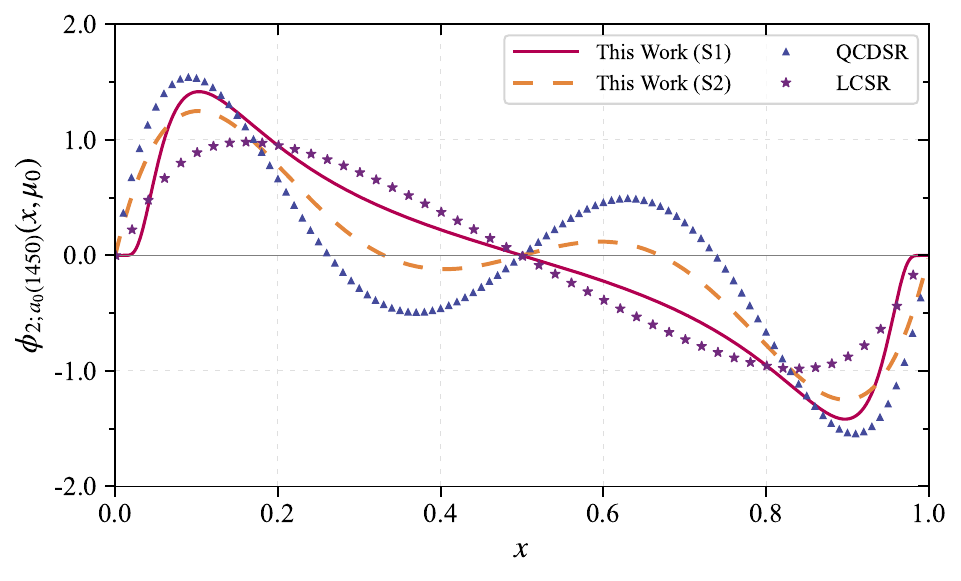}
\end{center}
\caption{Behavior of the two $a_0(1450)$-resonance twist-2 LCDA schemes at $\mu_0$ scale. As a comparison, the prediction results of QCDSR~\cite{Cheng:2005nb} and LCSR~\cite{Song:2025mfm} are also presented.}
\label{Fig:DA}
\end{figure}
The $a_0(1450)$-resonance twist-2 LCDAs within two different schemes are shown in Fig.~\ref{Fig:DA}, where the predictions from the QCDSR and LCSR are also given for comparison. In the isospin limit, all distribution amplitudes exhibit antisymmetric behavior. Both $\phi_{2;a_0(1450)}^{\rm(S2)}(x,\mu)$ scheme and the QCDSR analysis construct the distribution amplitude using the same Gegenbauer polynomial expansion, so their behaviors are consistent and differ only in the magnitude of the peak: the peak of QCDSR is located at $x=0.091$ with a value of $1.544$, while the peak of $\phi_{2;a_0(1450)}^{\rm(S2)}(x,\mu)$ is at $x=0.102$ with a value of $1.251$. In contrast, both $\phi_{2;a_0(1450)}^{\rm(S1)}(x,\mu)$ and the LCSR result are constructed within the LCHO model, but their free parameters are determined in different ways. In this work, the free parameters of $\phi_{2;a_0(1450)}^{\rm(S1)}(x,\mu)$ are fixed by the five $\xi$-moments, whereas the LCSR determines its free parameters via several constraint conditions. The peak value of LCSR result~\cite{Song:2025mfm} is $0.982$ at $x=0.159$, which is smaller than the $\phi_{2;a_0(1450)}^{\rm(S1)}(x,\mu)$ peak value of $1.418$ at $x=0.103$.

At the scale $\mu_k$, we take the Gegenbauer moments for the twist-3 two-particle LCDAs $\phi_{3;a_0(1450)}^p(x,\mu)$ and $\phi_{3;a_0(1450)}^\sigma(x,\mu)$ of the $a_0(1450)$-resonance as follows~\cite{Han:2013zg}:
\begin{align}
& a_2^p(\mu_k)=0.163\pm{0.005}, ~~~~~~ a_4^p(\mu_k)=0.316\pm{0.099},
\nonumber \\
& a_2^{\sigma}(\mu_k)=0.006\pm{0.001}, ~~~~~~ a_4^{\sigma}(\mu_k)=0.027\pm{0.005}.
\end{align}

\begin{table}[t]
\footnotesize
\begin{center}
   \caption{Numerical results of TFFs for the $\bar{B}^0\to a_0(1450)^+$ transition at the large recoil point. These results are calculated under two twist-2 LCDA schemes for $a_0(1450)$. For comparison, we also list the numerical results of LCSR~\cite{Han:2023pgf, Han:2013zg, Wang:2008da, Sun:2010nv} and pQCD~\cite{Li:2008tk}.}
  \label{The TFFs value}
  \centering
  \renewcommand{\arraystretch}{1}
  \begin{tabular}{l l l l}
    \hline
     ~~~~~~~~~~~~~~~& $f_+(0)$ ~~~~~~~~~~~~~~~& $f_-(0)$ ~~~~~~~~~~~~~~~& $f_{\rm T}(0)$ \\
    \hline
       This work ($\mathrm{S1}$) & $0.404_{-0.054}^{+0.048}$ & $-0.403_{-0.052}^{+0.047}$ & $0.501_{-0.060}^{+0.067}$ \\
    This work ($\mathrm{S2}$) & $0.398_{-0.053}^{+0.047}$ &
    $-0.404_{-0.048}^{+0.053}$ & $0.516_{-0.062}^{+0.069}$ \\
       LCSR~\cite{Han:2023pgf} & $0.40^{+0.01}_{-0.01}$ & $-0.39^{+0.01}_{-0.01}$ & $0.54^{+0.13}_{-0.13}$ \\
       LCSR~\cite{Han:2013zg} & $0.44_{-0.05}^{+0.06}$ & $-0.26_{-0.05}^{+0.06}$ & $0.43^{+0.06}_{-0.05}$ \\
    LCSR~\cite{Wang:2008da} & $0.52$ & $-0.44$ & $0.66$ \\
    LCSR~\cite{Sun:2010nv} & $0.53$ & $-0.53$ & $--$ \\
    pQCD~\cite{Li:2008tk} & $0.68_{-0.15}^{+0.19}$ & $--$ ~~~~~~~& $0.92^{+0.30}_{-0.21}$ \\
    \hline
  \end{tabular}
\end{center}
\end{table}
\begin{figure}[t]
\begin{center}
\includegraphics[width=0.45\textwidth]{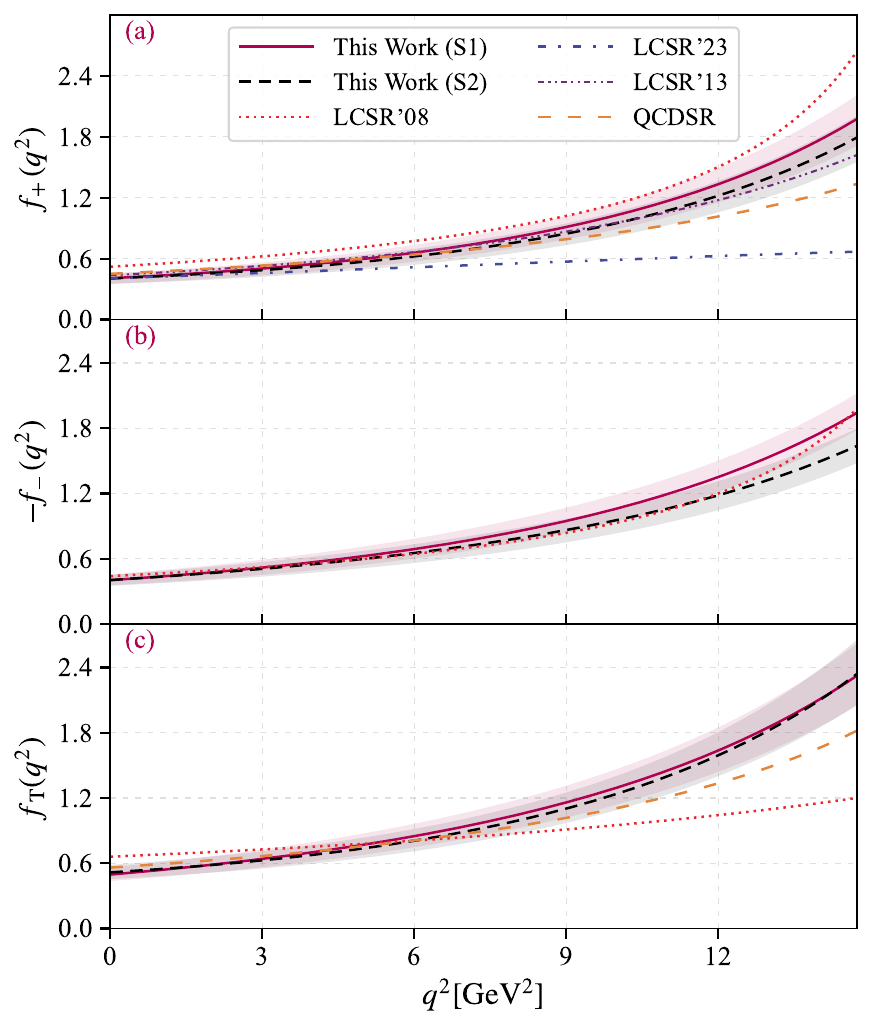}
\end{center}
\caption{Behaviors of $\bar{B}^0 \to a_0(1450)^+$ TFFs $f^{\rm{(S1,S2)}}_{\pm,{\rm T}}(q^2)$ in the entire $q^2$-region, where the solid lines represent the central values and the shaded regions represent the uncertainty ranges. For comparison, predictions from other theoretical groups are also provided.}
\label{Fig:TFFs}
\end{figure}
Subsequently, for the continuum threshold $s_0$ and the Borel parameter $M^2$ in the sum rules, we require that (i) the contributions from the continuum and higher excited states be less than $30\%$ and (ii) the dependence of the TFFs on $M^2$ be weak. Accordingly, we adopt: $s_0^{(+)}=22 \pm 0.1~{\rm GeV}^2$, $M^2_{(+)}=26\pm{1}~\rm{GeV}^2$, $s_0^{\rm{(-,T)}}=21\pm{0.1}~\rm{GeV}^2$, $M^2_{\rm{(-,T)}}=26\pm{1}~\rm{GeV}^2$. After considering the twist-2 and twist-3 contributions to the TFFs, we obtain the TFF results at $q^2 =0$, as listed in Table~\ref{The TFFs value}. It can be seen that although the two constructed LCDAs have different forms, the resulting TFFs are consistent with each other within uncertainties. In particular, the central values of $f_-^{{\rm (S1,S2)}}(0)$ differ by only 0.001. In addition, our $f_+^{\rm (S1,S2)}(q^2)$ agrees closely with the result of LCSR~\cite{Han:2023pgf}, while it is smaller than the results from LCSRs~\cite{Wang:2008da, Sun:2010nv} and pQCD~\cite{Li:2008tk}. For $f_-^{{\rm (S1,S2)}}(0)$, our results show better agreement with the LCSRs~\cite{Han:2023pgf, Wang:2008da}, whereas $f_{\rm T}^{{\rm (S1,S2)}}(0)$ displays a sizable discrepancy relative to the pQCD~\cite{Li:2008tk}.

Since the LCSR approach is mainly applicable in the low and intermediate $q^2$-region, we adopt the simplified series expansion (SSE) method to extrapolate $f_{\pm,\rm {T}}^{\rm(S1,S2)}(q^2)$ to the full physical $q^2$-region~\cite{Bourrely:2008za}. Fig.~\ref{Fig:TFFs} shows the $q^2$ dependence of the TFFs obtained from the two LCDA schemes, where the solid lines stand for the central values and the shaded bands denote the corresponding uncertainties. One can observe that the results from the two schemes are largely consistent over the entire $q^2$-region, with only slight deviations in the large $q^2$-region, indicating that the obtained TFFs exhibit good stability with respect to the parametrization of the twist-2 LCDA. In addition, we also present the predictions from the LCSR~\cite{Wang:2008da, Han:2013zg, Han:2023pgf} and QCDSR~\cite{Wang:2010dp} theory groups for comparison. One can see that $f_+^{\rm(S1,S2)}(q^2)$, $f_-^{\rm(S1,S2)}(q^2)$ and $f_{\rm T}^{\rm(S1,S2)}(q^2)$ all exhibit stable behaviors with increasing $q^2$ over the whole $q^2$-region.
\begin{figure}
\begin{center}
\includegraphics[width=0.45\textwidth]{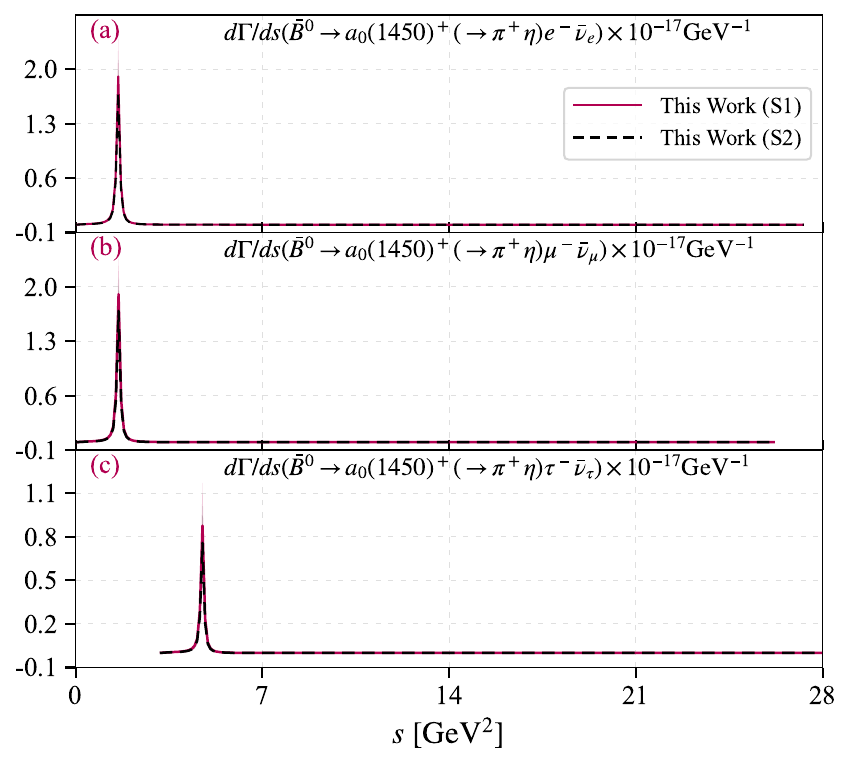}
\caption{Different decay widths for the quasi-four-body semileptonic decays $\bar{B}^0 \to a_0(1450)^+ (\to \pi^+ \eta) \ell^- \bar{\nu}_\ell$ with $\ell=(e,\mu,\tau)$, as functions of the $\pi\eta$ invariant-mass squared $s$. }
\label{Fig: decay width1}
\end{center}
\end{figure}
\begin{figure}
\begin{center}
\includegraphics[width=0.435\textwidth]{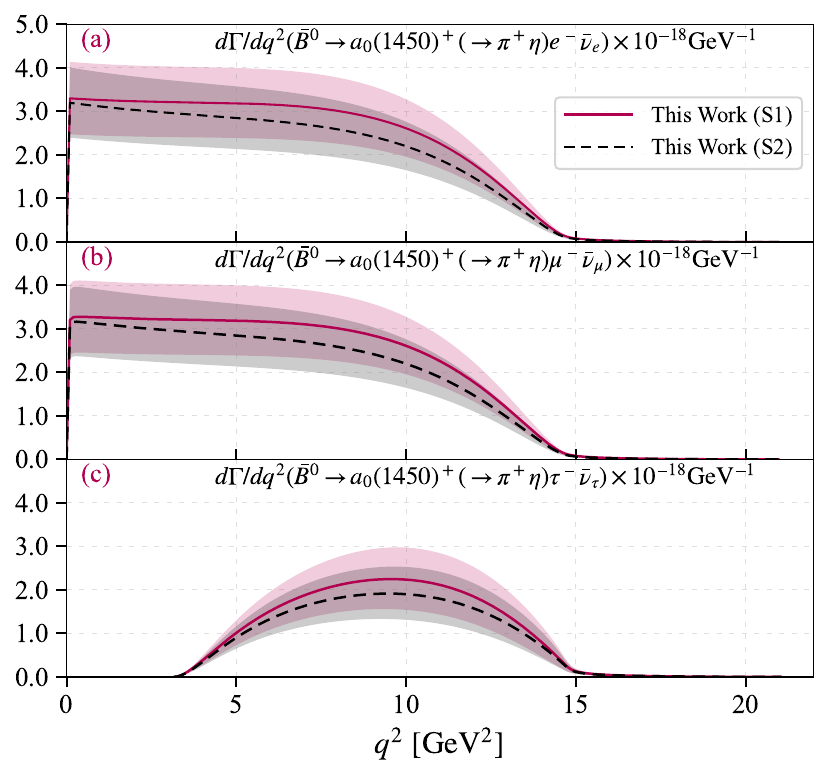}
\caption{Different decay widths for the quasi-four-body semileptonic decays $\bar{B}^0 \to a_0(1450)^+ (\to \pi^+ \eta) \ell^- \bar{\nu}_\ell$ with $\ell=(e,\mu,\tau)$, as functions of the lepton-antineutrino invariant-mass squared $q^2$.}
\label{Fig: decay width2}
\end{center}
\end{figure}

For the $\bar{B}^0 \to a_0(1450)^+ (\to \pi^+ \eta) \ell^- \bar{\nu}_\ell$ quasi-four-body semileptonic decay, in addition to the TFFs, the width effect of the intermediate $a_0(1450)$-resonance should also be considered. In practical calculations, we adopt $G_F =1.166 \times 10^{-5}~\rm{GeV^{-2}}$, $|V_{ub}|=3.82 \times 10^{-3}$, as well as $m_e=0.511~\mathrm{MeV}$, $m_\mu=105.659~\mathrm{MeV}$, and $m_\tau=1.77693~\mathrm{GeV}$. We take $\Gamma_{a_0(1450)}=258~\rm{MeV}$, and the corresponding strong coupling constants are determined using $\mathcal{B}(a_0(1450)\to \pi \eta)=0.093$ and $\mathcal{B}(a_0(1450)\to K \bar{K})=0.082$~\cite{ParticleDataGroup:2024cfk}. Figs.~\ref{Fig: decay width1}(a)--(c) display the $d\Gamma/ds$ distributions for the $e$, $\mu$, and $\tau$ channels, respectively. All three lepton channels exhibit prominent peaks in the $a_0(1450)$-resonance region around $s\simeq m_{a_0(1450)}^2$, while the energy-dependent phase-space factors of the $\pi\eta$ and $K\bar{K}$ channels further govern the finite width and line shape of the resonance. The spectra of the $e$ and $\mu$ channels are nearly identical, whereas the $\tau$-channel is substantially suppressed due to the large lepton mass and more restricted phase space. Moreover, the $s$ distributions obtained from the two LCDA schemes share the same resonance structure, but the central values of the first LCDA scheme are overall slightly higher. Similarly, Fig.~\ref{Fig: decay width2} shows the $d\Gamma/dq^2$ distributions. For the $e$ and $\mu$ channels, the differential width is large in the low $q^2$ region and gradually decreases with increasing $q^2$, becoming strongly suppressed for $q^2 \gtrsim 15~\rm{GeV}^2$. It should be noted that although the formal kinematic upper bound of $q^2$ for the quasi-four-body decay reads $q^2=(m_{\bar{B}^0}-(m_\pi +m_\eta))^2$, large $q^2$ only permits small values of the $\pi\eta$ invariant mass $s$, which progressively excludes the dominant resonance region around $s\simeq m_{a_0(1450)}^2$. As a result, the combined effect of the Flatt\'e spectral function and the quasi-four-body phase space makes the decay rate approach zero well before the formal $q^2$ endpoint. This is an important feature that distinguishes this quasi-four-body final-state decay from the three-body semileptonic decay in which the $a_0(1450)$-resonance is treated as an on-shell particle. For the $\tau$ channel, the nonzero lower bound of $q^2$ and the stronger phase-space compression lead to a distinct single-peak structure, with the peak located in the intermediate $q^2$-region.

\begin{table}[t]
\footnotesize
\begin{center}
\caption{Branching fractions (in unit: $10^{-4}$) for the quasi-four-body $\bar{B}^0 \to a_0(1450)^+ (\to \pi \eta)\ell^- \bar{\nu}_\ell$ and three-body NWA $\bar{B}^0 \to a_0(1450)^+ \ell^- \bar{\nu}_\ell$ semileptonic decays with $\ell=(e,\mu, \tau)$, respectively. For comparison, we also present the predictions of LCSR'13~\cite{Han:2013zg}.}
\label{table:Br}
\begin{tabular}{l l l l}
\hline
Decay Modes~~~~~~~~~~~~~~~~~~~~~~~~~ & This work (S1) & This work (S2) & LCSR'13 \\
\hline
 $\bar{B}^0\to a_0(1450)^+ (\to \pi^+ \eta) e^- \bar{\nu}_{e}$ & $0.883^{+0.224}_{-0.221}$ & $0.783^{+0.199}_{-0.196}$ & $0.892^{+0.260}_{-0.191}$ \\
 $\bar{B}^0\to a_0(1450)^+(\to \pi^+ \eta)\mu^- \bar{\nu}_\mu$ & $0.881^{+0.225}_{-0.221}$ & $0.781^{+0.199}_{-0.196}$ &
   $0.891^{+0.262}_{-0.191}$ \\
 $\bar{B}^0\to a_0(1450)^+(\to\pi^+\eta)\tau^- \bar{\nu}_{\tau}$ & $0.408^{+0.164}_{-0.145}$ & $0.353^{+0.144}_{-0.124}$ &
   $0.625^{+0.182}_{-0.134}$\\
 $\bar{B}^0\to a_0(1450)^+ e^- \bar{\nu}_{e}$ & $1.694^{+0.428}_{-0.423}$ & $1.504^{+0.383}_{-0.377}$ &
  $1.716^{+0.501}_{-0.368}$ \\
 $\bar{B}^0\to a_0(1450)^+ \mu^- \bar{\nu}_{\mu}$ & $1.691^{+0.429}_{-0.423}$ & $1.501^{+0.383}_{-0.377}$ &
  $1.715^{+0.500}_{-0.369}$ \\
 $\bar{B}^0\to a_0(1450)^+ \tau^- \bar{\nu}_{\tau}$ & $0.780^{+0.337}_{-0.276}$ & $0.673^{+0.273}_{-0.237}$ &
  $1.195^{+0.348}_{-0.256}$\\
\hline
\end{tabular}
\end{center}
\end{table}

By further integrating the double-differential decay width over the full physical phase space and adopting $\tau_{\bar{B}^0}=1.517~\rm{ps}$~\cite{ParticleDataGroup:2024cfk}, we obtain the branching fractions listed in Table~\ref{table:Br}. The results from the two LCDA schemes are consistent within their uncertainties, with the central values obtained in S1 slightly larger than those in S2, in agreement with the relative magnitudes of the differential decay widths shown in Figs.~\ref{Fig: decay width1} and \ref{Fig: decay width2}. The branching fractions of the electron and muon channels are nearly equal, indicating negligible lepton-mass effects in these two channels. In contrast, the branching fraction of the $\tau$ channel is considerably smaller, mainly because the larger $\tau$ lepton mass reduces the available phase space. Since no direct theoretical predictions are currently available for the quasi-four-body semileptonic decays $\bar{B}^0 \to a_0(1450)^+ (\to\pi^+\eta) \ell^- \bar{\nu}_\ell$ with $\ell=e,\mu,\tau$, we further employ the $\bar{B}^0 \to a_0(1450)^+$ TFFs obtained in LCSR'13~\cite{Han:2013zg} as inputs and evaluate the corresponding branching fractions within the same quasi-four-body decay framework in this work. The resulting predictions are denoted by LCSR'13 in Table~\ref{table:Br}. For the light-lepton channels, the LCSR'13 results are close to those obtained with S1 and are slightly larger than the S2 predictions. A more pronounced difference is found in the $\tau$ channel, where the LCSR'13 result is larger than those obtained with both S1 and S2. This comparison indicates that the $\tau$ channel exhibits a stronger sensitivity to the TFF inputs, particularly to their $q^2$ dependence. Table~\ref{table:Br} also presents the predictions for the $\bar{B}^0 \to a_0(1450)^+ \ell^- \bar{\nu}_\ell$ three-body semileptonic decay, where $ a_0(1450)$ is treated as a final state particle, for reference. These values are mainly used to illustrate the typical range of predictions for the weak transition $\bar{B}^0 \to a_0(1450)^+$ within different theoretical frameworks. Our quasi-four-body calculation explicitly incorporates the strong decay $a_0(1450)\to \pi \eta$, finite-width effects, and the integration over the $\pi \eta$ invariant mass, yielding theoretical predictions that correspond more directly to the experimental final state $\pi^+ \eta \ell^- \bar{\nu}_\ell$.

\mysection{Summary}\label {Sec:IV}
The internal structure of light scalar mesons remains highly controversial. Adopting the $a_0(1450)$-resonance as the lowest-Fock-state quark-antiquark component, we have investigated the $\bar{B}^0 \to a_0(1450)^+ (\to \pi^+ \eta) \ell^- \bar{\nu}_\ell$ quasi-four-body semileptonic decays with $\ell=(e,\mu,\tau)$. First, we calculated the first five odd $\xi$-moments $n=(1,3,5,7,9)$ of $a_0(1450)$-resonance twist-2 LCDA using QCDSR within the BFT. Based on these $\xi$-moments, we constructed two twist-2 LCDA schemes: one is the LCHO model (S1), whose parameters were determined by a simultaneous fit to the five odd moments, and the other is the truncated Gegenbauer expansion (S2) retaining only $a_1(\mu)$ and $a_3(\mu)$.

Subsequently, we substitute the two LCDA schemes separately as nonperturbative inputs into the LCSR to obtain the $\bar{B}^0 \to a_0(1450)^+$ TFFs $f_{\pm,\rm{T}}(q^2)$. The TFFs extracted from the two schemes are consistent within uncertainties and exhibit similar trends over the full physical $q^2$-region, in general agreement with existing theoretical results. On this basis, we employ the Flatt\'e line shape incorporating the coupled $\pi \eta$ and $K \bar{K}$ channels to describe the finite-width effect of $a_0(1450)$, and evaluate the double-differential decay widths and branching fractions for the quasi-four-body final state. The distribution $d\Gamma/ds$ shows a prominent resonance peak near $s \simeq m_{a_0(1450)}^2$, while $d\Gamma/dq^2$ is substantially suppressed for $q^2 \gtrsim (m_{\bar{B}^0}-m_{a_0(1450)})^2$. For the $e$ and $\mu$ channels, the two distributions nearly overlap, whereas the $\tau$-channel suffers evident phase-space suppression owing to the large lepton mass.

Finally, the quasi-four-body branching fractions obtained from the two LCDA schemes show good consistency within theoretical uncertainties, with only slightly larger central values for $\rm{S1}$ compared with $\rm{S2}$. We expect these results to provide a useful theoretical reference for future experimental studies of $\bar{B}^0 \to a_0(1450)^+ (\to \pi^+ \eta) \ell^- \bar{\nu}_\ell$, and to help further test the theoretical description of the $\bar{B}^0 \to a_0(1450)^+$ transition and of the internal structure of the $a_0(1450)$.

\mysection{Acknowledgments}
We are grateful to Prof. Tao Zhong for helpful discussion. This work was supported in part by the National Natural Science Foundation of China under Grant No.12265010, the Project of Guizhou Provincial Department of Science and Technology under Grant No.MS[2025]219, No.CXTD[2025]030.

\mysection{Appendix}\label {sec:VI}
For the final sum rule for the $a_0(1450)$-resonance twist-2 LCDA moments, the terms of perturbative part, two-gluon, three-gluon, four-quark condensates in Eq.~\eqref{xi0xin} have the following forms

\begin{widetext}
\begin{align}
& {\rm Im}\, I^{\rm pert}(s) = \frac3{16\pi(n+1)(n+2)} \big\{m_u[2n+5+(-1)^n]- m_d [(-1)^n (2n+3) +1]\big\},
\\
& \hat{\cal B}_M \hat I_{\langle G^2\rangle}(M^2) = \frac{\langle \alpha_s G^2\rangle}{16\pi M^4}\bigg\{ \big((-1)^n m_1 - m_2\big) \, \theta(n-1) \, \bigg[4\,n(-1)^n\ln\frac{M^2}{\mu^2}  \, -  \psi_4(n)-4n(-1)^n\big(\psi^{(0)}(n+1)
+2\gamma_E\big) \bigg]+ (-1)^n \big[m_1\, \theta(n-2)
\nonumber \\
& \hspace{1.8cm}
+2m_2\big]\bigg[\frac{2}{n}(n+1)(1+n(-1)^n)
-(2n+1)\psi_3(n) \bigg]+2m_1 +m_2\theta(n-2)\bigg\},
\\
& \hat{\mathcal{B}}_M\hat{I}_{\langle G^3\rangle}(M^2) = \frac{\langle g_s^3 f G^3\rangle}{1152\pi^2 M^6}\big((-1)^n m_1 - m_2\big) \, \bigg\{\theta(n-3)  \frac{9}{n-1} \big[\big(4\psi_1(n)+5  (-1)^n\big) n^3 - 4  \big(\psi_1(n)+1+(-1)^n\big) \, n^2 -(2+(-1)^n)n -2\big]
\nonumber\\
& \hspace{1.8cm} + \theta(n-2) \, \big[18n\psi_3(n)-9\big[(21n^2-17n-2)(-1)^n+2\big]\big] + \theta(n-1)\bigg[126(-1)^n n^2 + (3-5(-1)^n)n + 36n (2n-1 ) (-1)^n
\nonumber\\
& \hspace{1.8cm} \times \bigg[ \ln \frac{M^2}{\mu^2} - \psi^{(0)} (n+1) - 2 \gamma_E \bigg] - 54(-1)^n \bigg] - 2 n (-1)^n \bigg\},
\\
& \hat{\mathcal{B}}_M \hat{I}_{\langle q^4\rangle}(M^2)
= \frac{(2+\kappa^2) \, \langle g_s^2 q\bar{q}\rangle^2}{11664\,\pi^2 M^6} \,
\Bigg\{\big((-1)^n\, m_1 -m_2 \big)
\bigg[32\,(-1)^n n-818\,(-1)^n -12\,\big(12\,(-1)^n n-7\,[1+(-1)^n] \big)\ln\frac{M^2}{\mu^2}
-\frac{99\,\theta(n-3)}{n-1}
\nonumber\\
&\hspace{1.8cm}
\times \Big[\big(4\psi_1(n)
\!+\!5(-1)^n \big)n^3 -4\big(\psi_1(n) +(-1)^n \!+1 \big)n^2 \!-\!(2\!+\!(-1)^n)n -2\Big]-\theta(n-2)\bigg[6\Big([-5\!+\!51
(-1)^n]n-3(1+(-1)^n) \Big)
\nonumber\\
&\hspace{1.8cm}
\times \psi_3(n)
-3\Big(693\,(-1)^n n^2 +[102-637\,(-1)^n]n + 36\,(-1)^n -10\Big) \bigg]
-\theta\,(n-1) \bigg[\frac{6}{n} \bigg( 448(\,-1)^n n^3 -[59+35\,(-1)^n]\,n^2
\nonumber\\
&\hspace{1.8cm}
+2[66(-1)^n n +5(-1)^n \!-\!51]n^2 \ln\frac{M^2}{\mu^2}-[51+61(-1)^n]n
 \!+\!6[1+(-1)^n]\bigg)
\! -\!12[-6n+5(-1)^n\!+\!5]\psi_4(n)\bigg]
\!-\!\big(\psi^{(0)}(n\!+\!1)
\nonumber\\
&\hspace{1.8cm}
+2\gamma_E \big)
\bigg[84(1+(-1)^n) \!-\!144(-1)^n n
-12n[66(-1)^n n +5(-1)^n \!-\!51]\theta(n-1) \bigg] \!+\!538 \bigg] \!-\!818m_1 \big[\psi^{(0)}(n+1)\!+\!2\gamma_E \!-\!1 \big]\Bigg\},
\end{align}
In addition, the digamma function $\psi(n)$ is given by:
\begin{align}
\psi_k(n)=\psi\left(\frac{n+k-2}{2}\right)-\psi\left(\frac{n+k-3}{2}\right)+(-1)^{n+k}\ln 4,\qquad k=(1,2,3,4)
\end{align}
Within the LCSR framework, the detail expressions of $\bar{B}^0 \to a_0(1450)^+$ TFFs are derived from the correlator Eq.~\eqref{correlator}, which have the following forms,
\begin{align}
f_+^{\rm(S1,S2)}(q^2) &=\frac{m_b \bar{f}_{a_0(1450)}}{2m_{\bar{B}^0}^2 f_{\bar{B}^0}} \int_{u_0}^{\tilde{u}_0} du e^{[um_{\bar{B}}^2-(m_b^2-\bar{u}q^2+u\bar{u}m_{a_0(1450)}^2)]/(uM^2)}
\Bigg\{-\frac{m_b}{u}\phi_{2;a_0(1450)}^{\rm(S1,S2)}(x,\mu)+m_{a_0(1450)}
\phi_{3;a_0(1450)}^p(x,\mu)
\nonumber \\
& +\frac{m_{a_0(1450)}}{6} \Bigg[\frac{2}{u}+\frac{4u m_b^2 m_{a_0(1450)}^2}{(m_b^2-q^2+u^2 m_{a_0(1450)}^2)^2}
-\frac{m_b^2+q^2-u^2 m_{a_0(1450)}^2}{m_b^2-q^2+u^2 m_{a_0(1450)}^2} \frac{d}{du}\Bigg]
\phi_{3;a_0(1450)}^\sigma(x,\mu)\Bigg\},
\\
\tilde{f}^{\rm(S1,S2)}(q^2) &=\frac{m_b \bar{f}_{a_0(1450)} m_{a_0(1450)}}{m_{\bar{B}^0}^2 f_{\bar{B}^0}}\int_{u_0}^{\tilde{u}_0} \frac{du}u  e^{[um_{\bar{B}}^2 - (m_b^2 -\bar{u}q^2+u\bar{u}m_{a_0(1450)}^2)]/(uM^2)} \Bigg[\phi_{3;a_0(1450)}^p (x, \mu) + \frac16 \frac{d}{du}\phi_{3;a_0(1450)}^\sigma(x,\mu) \Bigg],
\\
f_{\mathrm{T}}^{\rm(S1,S2)}(q^2) &=\frac{(m_{\bar{B}^0} \, +  \, m_{a_0(1450)}) \, m_b  \, \bar{f}_{a_0(1450)}} {m_{\bar{B}^0}^2 f_{\bar{B}^0}} \,
\int_{u_0}^{\tilde{u}_0}  \, du  \, e^{[um_{\bar{B}}^2 - (m_b^2-\bar{u}q^2+u\bar{u}m_{a_0(1450)}^2)]/(uM^2)}
\bigg\{ \, - \,  \frac{1}{2u}\phi_{2;a_0(1450)}^{\rm(S1,S2)}(x,\mu)  \, + \,  \frac16
\nonumber
\\
& \times\frac{m_b m_{a_0(1450)}}{m_b^2 -q^2 + u^2 m_{a_0(1450)}^2}\bigg(\frac{2u m_{a_0(1450)}^2}{m_b^2 -q^2 +u^2 m_{a_0(1450)}^2}- \frac{d}{du}\bigg)\phi_{3;a_0(1450)}^\sigma(x,\mu)\bigg\},
\label{formula TFFs}
\end{align}
where $s_0$ is the continuum threshold parameter, the lower and upper integration limits are given by:

\begin{align}
u_0(\tilde{u}_0) = \frac1{2m_{a_0(1450)}^2} \big\{[(q^2-s_0(m_b ^2)+m_{a_0(1450)}^2)^2 +4m_{a_0(1450)}^2
(m_b ^2-q^2)]^{1/2} +q^2-s_0(m_b ^2) + m_{a_0(1450)}^2 \big\},
\end{align}
\end{widetext}


\begin{thebibliography}{99}

\bibitem{ParticleDataGroup:2026mpi}
F.~Takahashi \textit{et al.} [Particle Data Group],
\href{https://doi.org/10.1142/S0217751X26300115}
{Int. J. Mod. Phys. A \textbf{41} (2026) 2630011}.



\bibitem{CrystalBarrel:1994arw}
Amsler, C. and others [Crystal Barrel Collaboration],
\href{https://doi.org/10.1016/0370-2693(94)91044-8}
{Phys. Lett. B \textbf{333} (1994) 277}.

\bibitem{Bugg:2008ig}
D.~V.~Bugg,
\href{https://doi.org/10.1103/PhysRevD.78.074023}
{Phys. Rev. D \textbf{78} (2008) 074023}.

\bibitem{BaBar:2007yhw}
B.~Aubert \textit{et al.} [BaBar Collaboration],
\href{https://doi.org/10.1103/PhysRevD.75.111102}
{Phys. Rev. D \textbf{75} (2007) 111102}.

\bibitem{Brito:2004tv}
T.~V.~Brito, F.~S.~Navarra, M.~Nielsen and M.~E.~Bracco,
\href{https://doi.org/10.1016/j.physletb.2005.01.008}
{Phys. Lett. B \textbf{608} (2005) 69}.


\bibitem{Klempt:2007cp}
E.~Klempt and A.~Zaitsev,
\href{https://doi.org/10.1016/j.physrep.2007.07.006}
{Phys. Rept. \textbf{454} (2007) 1}.

\bibitem{Balitsky:1989ry}
I.~I.~Balitsky, V.~M.~Braun and A.~V.~Kolesnichenko,
\href{https://doi.org/10.1016/0550-3213(89)90570-1}
{Nucl. Phys. B \textbf{312} (1989) 509}.

\bibitem{Chernyak:1990ag}
V.~L.~Chernyak and I.~R.~Zhitnitsky,
\href{https://doi.org/10.1016/0550-3213(90)90612-H}
{Nucl. Phys. B \textbf{345} (1990) 137}.

\bibitem{Wang:2008da}
Y.~M.~Wang, M.~J.~Aslam and C.~D.~L\"u,
\href{https://doi.org/10.1103/PhysRevD.78.014006}
{Phys. Rev. D \textbf{78} (2008) 014006.}


\bibitem{Han:2013zg}
H.~Y.~Han, X.~G.~Wu, H.~B.~Fu, Q.~L.~Zhang and T.~Zhong,
\href{https://doi.org/10.1140/epja/i2013-13078-7}
{Eur. Phys. J. A \textbf{49} (2013) 78.}

\bibitem{Galkin:2025emi}
V.~O.~Galkin and I.~S.~Sukhanov,
\href{https://doi:10.1103/PhysRevD.111.093001}
{Phys. Rev. D \textbf{111} (2025) 093001}.

\bibitem{Huang:2021owr}
Q.~Huang, Y.~J.~Sun, D.~Gao, G.~H.~Zhao, B.~Wang and W.~Hong,
\href{https://arxiv.org/abs/2102.12241}
{arXiv:2102.12241}.

\bibitem{Han:2023pgf}
X.~Y.~Han, L.~S.~Lu, C.~D.~L{\"u}, Y.~L.~Shen and B.~X.~Shi,
\href{https://doi.org/10.1007/JHEP11(2023)091}
{JHEP \textbf{11} (2023) 091}.



\bibitem{Song:2025mfm}
Y.~L.~Song, Y.~L.~Yang, Y.~Cao, X.~Zheng and H.~B.~Fu,
\href{https://doi.org/10.1088/1674-1137/ae5ef7}
{Chin. Phys. C \textbf{50} (2026) 073103}.

\bibitem{Verma:2011yw}
R.~C.~Verma,
\href{https://doi:10.1088/0954-3899/39/2/025005}
{J. Phys. G \textbf{39} (2012) 025005}.

\bibitem{Shifman:2001ck}
M.~Shifman and B.~Ioffe,
World Scientific, 2001,
\href{https://doi.org/10.1142/4544}
{ISBN 978-981-02-4445-3, 978-981-4492-22-5}.


\bibitem{Khodjamirian:2000ds}
A.~Khodjamirian, R.~Ruckl, S.~Weinzierl, C.~W.~Winhart and O.~I.~Yakovlev,
\href{https://doi.org/10.1103/PhysRevD.62.114002}
{Phys. Rev. D \textbf{62} (2000) 114002}.

\bibitem{Duplancic:2008ix}
G.~Duplancic, A.~Khodjamirian, T.~Mannel, B.~Melic and N.~Offen,
\href{https://doi.org/10.1088/1126-6708/2008/04/014}
{JHEP \textbf{04} (2008) 014}.

\bibitem{Wang:2007fs}
Y.~M.~Wang and C.~D.~Lu,
\href{https://doi.org/10.1103/PhysRevD.77.054003}
{Phys. Rev. D \textbf{77} (2008), 054003}.

\bibitem{Ali:1993vd}
A.~Ali, V.~M.~Braun and H.~Simma,
\href{https://doi.org/10.1007/BF01580324}
{Z. Phys. C \textbf{63} (1994) 437-454}.

\bibitem{Aliev:1995zlh}
T.~M.~Aliev, D.~A.~Demir, E.~Iltan and N.~K.~Pak,
\href{https://doi.org/10.1103/PhysRevD.54.857}
{Phys. Rev. D \textbf{54} (1996) 857-862}.

\bibitem{Khodjamirian:2000mi}
A.~Khodjamirian,
\href{https://doi.org/10.1016/S0550-3213(01)00194-8}
{Nucl. Phys. B \textbf{605} (2001) 558-578}.

\bibitem{Khodjamirian:2002pk}
A.~Khodjamirian, T.~Mannel and P.~Urban,
\href{https://doi.org/10.1103/PhysRevD.67.054027}
{Phys. Rev. D \textbf{67} (2003) 054027}.

\bibitem{Khodjamirian:2003eq}
A.~Khodjamirian, T.~Mannel and B.~Melic,
\href{https://doi.org/10.1016/j.physletb.2003.08.012}
{Phys. Lett. B \textbf{571} (2003) 75-84}.

\bibitem{Khodjamirian:2005wn}
A.~Khodjamirian, T.~Mannel, M.~Melcher and B.~Melic,
\href{https://doi.org/10.1103/PhysRevD.72.094012}
{Phys. Rev. D \textbf{72} (2005) 094012}.

\bibitem{Belyaev:1994zk}
V.~M.~Belyaev, V.~M.~Braun, A.~Khodjamirian and R.~Ruckl,
\href{https://doi.org/10.1103/PhysRevD.51.6177}
{Phys. Rev. D \textbf{51} (1995) 6177-6195}.

\bibitem{Zhong:2022lmn}
T.~Zhong, Z.~H.~Zhu and H.~B.~Fu,
\href{https://doi.org/10.1088/1674-1137/ac9deb}
{Chin. Phys. C \textbf{47} (2023) 013111}.


\bibitem{Huang:2022xny}
D.~Huang, T.~Zhong, H.~B.~Fu, Z.~H.~Wu, X.~G.~Wu and H.~Tong,
\href{https://doi.org/10.1140/epjc/s10052-023-11851-x}
{Eur. Phys. J. C \textbf{83} (2023) 680}.


\bibitem{Hu:2023pdl}
D.~D.~Hu, X.~G.~Wu, H.~B.~Fu, T.~Zhong, Z.~H.~Wu and L.~Zeng,
\href{https://doi.org/10.1140/epjc/s10052-023-12333-w}
{Eur. Phys. J. C \textbf{84} (2024) 15}.


\bibitem{Hu:2021lkl}
D.~D.~Hu, H.~B.~Fu, T.~Zhong, Z.~H.~Wu and X.~G.~Wu,
\href{https://doi.org/10.1140/epjc/s10052-022-10555-y}
{Eur. Phys. J. C \textbf{82} (2022) 603}.


\bibitem{Zhong:2014fma}
T.~Zhong, X.~G.~Wu and T.~Huang,
\href{https://doi.org/10.1140/epjc/s10052-015-3271-6}
{Eur. Phys. J. C \textbf{75} (2015) 45}.

\bibitem{Zhong:2014jla}
T.~Zhong, X.~G.~Wu, Z.~G.~Wang, T.~Huang, H.~B.~Fu and H.~Y.~Han,
\href{https://doi.org/10.1103/PhysRevD.90.016004}
{Phys. Rev. D \textbf{90} (2014) 016004.}


\bibitem{Huang:1994dy}
T.~Huang, B.~Q.~Ma and Q.~X.~Shen,
\href{https://doi.org/10.1103/PhysRevD.49.1490}
{Phys. Rev. D \textbf{49} (1994) 1490-1499}.

\bibitem{Yang:2005bv}
M.~Z.~Yang,
\href{https://doi.org/10.1103/PhysRevD.73.079901}
{Phys. Rev. D \textbf{73} (2006) 034027}.


\bibitem{Cheng:2005nb}
H.~Y.~Cheng, C.~K.~Chua and K.~C.~Yang
\href{https://doi.org/10.1103/PhysRevD.73.014017}
{Phys. Rev. D \textbf{73} (2006) 014017.}

\bibitem{Ball:2005vx}
P.~Ball and R.~Zwicky,
\href{https://doi.org/10.1016/j.physletb.2005.11.068}
{Phys. Lett. B \textbf{633} (2006) 289-297}.

\bibitem{Lu:2006fr}
C.~D.~L\"u, Y.~M.~Wang and H.~Zou,
\href{https://doi.org/10.1103/PhysRevD.75.056001}
{Phys. Rev. D \textbf{75} (2007) 056001}.

\bibitem{Zhong:2021epq}
T.~Zhong, Z.~H.~Zhu, H.~B.~Fu, X.~G.~Wu and T.~Huang,
\href{https://doi.org/10.1103/PhysRevD.104.016021}
{Phys. Rev. D \textbf{104} (2021) 016021.}

\bibitem{Zhong:2022ecl}
T.~Zhong, H.~B.~Fu and X.~G.~Wu,
\href{https://doi.org/10.1103/PhysRevD.105.116020}
{Phys. Rev. D \textbf{105} (2022) 116020}.



\bibitem{Wang:2016wpc}
W.~Wang,
\href{https://doi.org/10.1016/j.physletb.2016.06.007}
{Phys. Lett. B \textbf{759} (2016) 501}.

\bibitem{Achasov:2005hm}
N.~N.~Achasov and A.~V.~Kiselev,
\href{https://doi.org/10.1103/PhysRevD.73.054029}
{Phys. Rev. D \textbf{73} (2006) 054029}.

\bibitem{Dosch:2002rh}
H.~G.~Dosch, E.~M.~Ferreira, F.~S.~Navarra and M.~Nielsen,
\href{https://doi.org/10.1103/PhysRevD.65.114002}
{Phys. Rev. D \textbf{65} (2002) 114002}.

\bibitem{ParticleDataGroup:2024cfk}
S.~Navas \textit{et al.} [Particle Data Group],
\href{https://doi.org/10.1103/PhysRevD.110.030001}
{Phys. Rev. D \textbf{110} (2024) 030001.}

\bibitem{Gelhausen:2013wia}
P.~Gelhausen, A.~Khodjamirian, A.~A.~Pivovarov and D.~Rosenthal,
\href{https://doi.org/10.1103/PhysRevD.88.014015}
{Phys. Rev. D \textbf{88} (2013) 014015.}

\bibitem{Colangelo:2000dp}
P.~Colangelo and A.~Khodjamirian,
\href{https://arxiv.org/abs/hep-ph/0010175}
{hep-ph/0010175}.

\bibitem{Sun:2010nv}
Y.~J.~Sun, Z.~H.~Li and T.~Huang,
\href{https://doi.org/10.1103/PhysRevD.83.025024}
{Phys. Rev. D \textbf{83} (2011) 025024}.

\bibitem{Li:2008tk}
R.~H.~Li, C.~D.~L\"u, W.~Wang and X.~X.~Wang,
\href{https://doi.org/10.1103/PhysRevD.79.014013}
{Phys. Rev. D \textbf{79} (2009) 014013.}

\bibitem{Bourrely:2008za}
C.~Bourrely, I.~Caprini and L.~Lellouch,
\href{https://doi.org/10.1103/PhysRevD.82.099902}
{Phys. Rev. D \textbf{79} (2009) 013008}.

\bibitem{Wang:2010dp}
Z.~G.~Wang and J.~F.~Li,
\href{https://arxiv.org/abs/1012.1704}
{arXiv:1012.1704}.



\end{thebibliography}
\end{document}